\documentstyle[preprint,aps,eqsecnum]{revtex}

\newcommand{\tto}{{3{\mathchar"002D}3{\mathchar"002D}1}}
\begin{document}
\draft
\preprint{\vbox{
\hbox{CTP-TAMU-3/95}
\hbox{IFP-709-UNC}
\hbox{hep-ph/9502353}
\hbox{January 1995}
}}

\title{The decay $b\to s\gamma$ in the 3-3-1 model}

\author{Jyoti Agrawal, Paul H. Frampton}
\address{
Institute of Field Physics, Department of Physics and Astronomy,\\
University of North Carolina, Chapel Hill, NC  27599--3255, USA
}
\author{James T. Liu}
\address{
Center for Theoretical Physics, Department of Physics\\
Texas A\&M University, College Station, TX 77843--4242
}
\maketitle
\begin{abstract}
The 3-3-1 model, based on the gauge group $SU(3)_c\times SU(3)_L\times
U(1)_X$, makes a natural prediction of three generations based on anomaly
cancellation.  Since this is accomplished by incorporating the third family
of quarks differently from the other two, it leads to potentially large
flavor changing neutral currents.  A sensitive place to look for such
effects is the flavor changing $b\to s\gamma$ decay, which has recently
been measured at CLEO.  We compute this decay rate in the 3-3-1 model and
compare it with that of the two-Higgs-doublet model, a subset of the full
3-3-1 model.  We find that the additional 3-3-1 physics weakens the bound on
the charged Higgs mass from $M_{H^+}>290$~GeV to $M_{H^+}\agt120$~GeV.
\end{abstract}

\narrowtext

\section{Introduction}

Since most indirect effects of new physics first enter at the loop level,
they are often dominated by tree level Standard Model (SM) contributions.
It is for this reason that only now, with the advent of precision
electroweak measurements, we are beginning to probe the structure of new
physics.  However an exception to this approach is the process $b\to s
\gamma$.  Since this Flavor Changing Neutral Current (FCNC) process first
occurs at loop level in the SM, we have an interesting case where the
effects of new physics may be comparable to the SM contribution.

On the experimental side, CLEO has recently announced a measurement of the
inclusive decay rate \cite{cleo}
\begin{equation}
{\rm BR}(b\to s\gamma)=(2.32\pm0.51\pm0.29\pm0.32)\times 10^{-4}\ ,
\end{equation}
where the final two errors are both systematic in nature.  This translates
into
\begin{equation}
1\times 10^{-4} < {\rm BR}(b\to s\gamma) < 4\times 10^{-4}\ ,
\label{eq:cleo}
\end{equation}
at the 95\% confidence level.  While there are still substantial theoretical
uncertainties in the SM prediction for $b\to s\gamma$ (mostly related to
unknown next leading order QCD corrections \cite{buras}), this experimental
value agrees well with a standard model prediction of
$(2.75\pm0.80)\times10^{-4}$ with a top quark mass $m_t=175$~GeV
\cite{buras,cdf,grinstein,hewett}.

The sensitivity of this FCNC process to new physics thus allows us to put
limits on many theories of physics beyond the SM \cite{hewett}.  In this
paper we examine the implications of the $b\to s\gamma$ penguin on the 3-3-1
model \cite{ppframpton}.  This model, based on the gauge group
$SU(3)_c\times SU(3)_L\times U(1)_X$, gives a potential answer to the
question of flavor by predicting three families as a consequence of anomaly
cancellation.  This is accomplished by making one of the three quark
families transform differently from the other two under the $SU(3)_L\times
U(1)_X$ extended electroweak gauge group.  Hence this model suffers
generically from large FCNCs \cite{ppframpton,liu,ppfcnc}, leading to a
possible large enhancement of the $b\to s\gamma$ decay rate.

While the 3-3-1 model predicts no new leptons, it predicts one new quark per
family, denoted here by $D$ and $S$ with charge $-4/3$ and $T$ with charge
$5/3$ \cite{ppframpton}.  These quarks interact with the ordinary quarks via
the charged dilepton gauge boson doublet, $(Y^{++},Y^+)$.  In addition,
there is a neutral $Z'$ gauge boson with flavor changing couplings to the
usual quarks.  Thus there are
new processes contributing to $b\to s\gamma$ incorporating $Y$--new quark
loops as well as flavor changing $Z'$ loops.
Although many Higgs multiplets
are necessary in order to break the 3-3-1 gauge symmetry, we show that it is
only necessary to look at an effective two-Higgs-doublet model.

An unusual feature of the
3-3-1 model are the strong bounds placed on the masses of the new particles.
Of present interest is the bound $300<M_Y\alt1100$~GeV \cite{phenom,liung}
which indicates that the 3-3-1 contributions to $b\to s\gamma$ may not
be suppressed by simply increasing the 3-3-1 scale, $M_Y$.

As is well known, QCD corrections lead to an important enhancement of the
$b\to s\gamma$ decay rate in the SM.  Thus, when examining the additional
contribution from the 3-3-1 model, we divide the calculation into two parts.
In section 2, we calculate the new electroweak penguin diagrams associated
with the 3-3-1 model, and in section 3, we examine the effects of QCD
running from the 3-3-1 scale to $b$-quark scale.  In section 4 we combine
the results of the previous sections and examine the implications of the
current CLEO bound on the 3-3-1 model.  Finally we present our conclusions
in section 5.

\section{Electroweak penguins in the SM and 3-3-1 model}

Although the 3-3-1 model contains tree level FCNC interactions mediated
by $Z'$ exchange, the decay $b\to s\gamma$ remains a loop process in this
model.  Thus we expect the additional 3-3-1 contributions to $b\to s\gamma$
to be at most comparable to the SM penguin diagram.  These new 3-3-1
contributions arise from dilepton gauge boson $Y$, $Z'$ and charged Higgs
loops.  At one-electroweak-loop order, all these contributions add linearly
to the $b\to s\gamma$ amplitude.  Hence we consider them one at a time.

In the gauge sector all lowest order contributions to $b\to s\gamma$ are
given by the penguin diagrams of Fig.~\ref{fig:1} where $V=W$, $Y$ or $Z'$.
Since all charged gauge bosons contribute similarly, we first calculate the
effective $b\to s\gamma$ vertex for an arbitrary left handed gauge boson
loop and subsequently specialize to $W$ or $Y$ loops.  The $Z'$ contribution
is calculated separately since it involves flavor changing vertices.

For a left handed gauge boson $V$, the diagrams of Fig.~\ref{fig:1} give
rise to an effective vertex
\begin{equation}
\Gamma_\mu=e{\alpha\over4\pi s^2}{1\over2M_V^2}\overline{s}\left[
(q^2\gamma_\mu-q_\mu\!\not\!q)F_1(q^2)+(i\sigma_{\mu\nu}q^\nu\!\!\not\!p_1
+\!\not\!p_2i\sigma_{\mu\nu}q^\nu)F_2(q^2)+M_V^2\gamma_\mu F^{\rm nAb}(q^2)
\right]\gamma_Lb ,
\label{eq:veff}
\end{equation}
where $q_\mu$ is the momentum carried by the photon.  While gauge invariance
of the photon demands a vanishing $F^{\rm nAb}$, we will see that this arises
in a subtle manner in the 3-3-1 model.  We work in Feynman gauge and ignore
the light quark masses.  Then, for a single fermion of mass $m$ in the loop,
we find the on-shell form factors
\begin{eqnarray}
F_1(0)&=&Q\left[{1\over9}+{x(x^2+11x-18)\over12(x-1)^3}
+{-9x^2+16x-4\over6(x-1)^4}\ln x\right]\nonumber\\
&&+Q_V\left[-{8\over9}+{x(7x^2-x-12)\over12(x-1)^3}
+{x^2(x^2-10x+12)\over6(x-1)^4}\ln x\right]\nonumber\\
F_2(0)&=&Q\left[{2\over3}-f_1(x)\right]+Q_V\left[-{5\over6}-f_2(x)\right]\ ,
\label{eq:penguin}
\end{eqnarray}
where $x=m^2/M^2$.  For later convenience, we have defined the functions
\begin{eqnarray}
f_1(x)&=&{x(x^2-5x-2)\over4(x-1)^3}+{3\over2}{x^2\over(x-1)^4}\ln x
\nonumber\\
f_2(x)&=&{x(-2x^2-5x+1)\over4(x-1)^3}+{3\over2}{x^3\over(x-1)^4}\ln x\ .
\end{eqnarray}
In the above form factors, $Q$ is the charge of the
fermion in the loop and $Q_V$ is the charge of the gauge boson.  In the SM,
only the $W$ loop is present, in which case $Q=2/3$ and $Q_V=-1$.  After
summing over all three families, the constant terms in Eqn.~(\ref{eq:penguin})
vanish by the GIM mechanism while the non-constant terms agree with the
expressions given by Inami and Lim \cite{inami}.

Prior to imposing the GIM mechanism, the gauge non invariant term
$F^{\rm nAb}$ is present.  In Feynman gauge, to the same order as above, we
find
\begin{equation}
F^{\rm nAb}(q^2)=2Q_V\left[\Delta_M+{q^2\over6M^2}+\cdots\right]\ ,
\label{eq:nab}
\end{equation}
where $\Delta_M={1\over\epsilon}-\gamma-\ln{-M^2\over4\pi\mu^2}$ in
dimensional regularization.  We note that $F^{\rm nAb}$ is independent of
the mass $m$ of the quark propagating in the loop.  Thus, assuming
generation universality as in the case of the SM, it drops out of the
final expression because of the GIM mechanism.

When we take QCD corrections into account we need to calculate the induced
$b\to sg$ penguins as well.  The above form factors are also valid for the
gluon penguin
provided we set $Q=1$ and $Q_V=0$ and replace $e$ by the strong coupling
$g_3T^a$.  For on-shell interactions, $F_1$ is unimportant and we are left
with the dipole terms given by $F_2$.  By convention, we now denote the
coefficient of such photon and gluon dipole terms by $C_7$ and $C_8$
respectively \cite{grinstein}.  In the case of a $W$ loop, we take the GIM
mechanism into account and assume a heavy top quark.  In this case the photon
and gluon dipole coefficients become%
\footnote{In this section we include the CKM factors explicitly.  However
subsequently they will be absorbed into the definition of the effective
Hamiltonian.}
\begin{eqnarray}
C_7^{\rm SM}(M_W)&=&-{1\over2}V_{ts}^*V_{tb}^{\vphantom{*}}
\left[{2\over3}f_1(x_t)-f_2(x_t)\right]
\nonumber\\
C_8^{\rm SM}(M_W)&=&-{1\over2}V_{ts}^*V_{tb}^{\vphantom{*}}f_1(x_t)\ ,
\label{eq:wloop}
\end{eqnarray}
where $x_t=m_t^2/M_W^2$.  This is the only contribution in the SM.

In the 3-3-1 model, we need to include the contributions from the
dilepton--heavy quark loops.  In this case, there is a generalized GIM
mechanism which replaces the standard case in the SM.  Denoting the unitary
mixing matrix in the down quark sector by $V_L$ with elements $v_{ij}$
\cite{liu}, we sum over all three families (exotic quarks $D$, $S$ and $T$)
to find
\begin{eqnarray}
C_7^Y(M_Y)&=&-{1\over2}\sum_{i=1}^3v_{is}^*v_{ib}^{\vphantom{*}}\left[
-{4\over3}f_1(x_i)+f_2(x_i)\right]\nonumber\\
&&-{1\over2}v_{3s}^*v_{3b}^{\vphantom{*}}\left[-{9\over2}+3f_1(x_3)-3f_2(x_3)
\right]
\nonumber\\
C_8^Y(M_Y)&=&-{1\over2}\sum_{i=1}^3v_{is}^*v_{ib}^{\vphantom{*}}f_1(x_i)\ ,
\label{eq:yloop}
\end{eqnarray}
where $x_i=m_{Q_i}^2/M_Y^2$ and $Q_i=(D,S,T)$.
Unlike the SM contribution, where we ignore masses of the first two
families, in this case phenomenology dictates that all exotic quarks are
heavy.  The gluon
penguin has the same form as in the SM since QCD is insensitive to the
electric charges of the quarks and gauge bosons in the loop.  On the other
hand, the second line in Eqn.~(\ref{eq:yloop}) is present because the third
family couples differently.  This is already well known for the tree level
$Z'$ couplings.  However we see here that generation non-universality also
appears in dilepton loop diagrams.

We note that this imperfect GIM cancellation has a couple of consequences.
First of all, unlike the SM, the dilepton induced $b\to s\gamma$ penguin in
the 3-3-1 model may be non-vanishing even when all exotic quarks are
degenerate in mass.  This is simply another manifestation of potentially
large FCNCs in the 3-3-1 model.  Secondly, we may worry about the divergent
gauge non-invariant term $F^{\rm nAb}$ of Eqn.~(\ref{eq:nab}).  Because the GIM
sum involves the gauge bosons $Y^+$ and $Y^{--}$, we find a left over
divergent term of the form
$-3v_{3s}^*v_{3b}^{\vphantom{*}}[\Delta_M+q^2/6M^2]$.
This, however, is cancelled by the $Z'$ FCNC vertex induced by $\gamma$--$Z'$
mixing as shown in Fig.~\ref{fig:2} \cite{agrawal}.  Since this diagram does
not contribute to the dipole form factor, its only purpose for the present
discussion is to eliminate the unwanted $F^{\rm nAb}$, and can otherwise be
ignored.

Finally, the diagrams with a $Z'$ in the loop also contribute to $b\to
s\gamma$ due to the flavor changing $Z'$ couplings.  In this neutral current
case, since the quark in the loop is light, we ignore its mass and find
\begin{eqnarray}
C_7^{Z'}(M_{Z'})&=&-{1\over3}C_8^{Z'}(M_{Z'})\nonumber\\
C_8^{Z'}(M_{Z'})&=&-{1\over2}v_{3s}^*v_{3b}^{\vphantom{*}}{-16\over9}
{s^2\over1-4s^2}\ ,
\end{eqnarray}
where $s^2=1-c^2=\sin^2\theta_W$ arises from the $Z'$ coupling to quarks
\cite{liu,daniel}.
In principle, this introduces yet another scale into the problem, namely
$M_{Z'}$.  Since the
$Z'$ is considerably heavier than the dileptons, we would in principle need
to worry about all three heavy scales, $M_{Z'}$, $M_Y$ and $M_W$ (in
addition to heavy quark thresholds) when including QCD corrections.
However, as a simplification we ignore the difference between the two 3-3-1
scales since the QCD running effects are less pronounced at higher energies.
In this case, we find it convenient to rewrite the $Z'$ contribution to
change the mass scale in the effective vertex, Eqn.~(\ref{eq:veff}),  from
$M_{Z'}$ to $M_Y$.  Using the relation
$M_{Z'}^2=M_Y^2/\rho_\tto\sin^2\theta_\tto$ where
$\cos^2\theta_\tto=3\tan^2\theta_W$ \cite{liung}, we arrive at
\begin{eqnarray}
C_7^{Z'}(M_Y)&=&-{1\over3}C_8^{Z'}(M_Y)\nonumber\\
C_8^{Z'}(M_Y)&=&-{1\over2}v_{3s}^*v_{3b}^{\vphantom{*}}{-16\over9}
\rho_\tto {s^2\over c^2}\ .
\label{eq:zploop}
\end{eqnarray}
As in the SM, the generalized rho parameter, $\rho_\tto$, depends on the
specifics of the extended Higgs sector.  $\rho_\tto=3/4$ in the minimal
3-3-1 model where $SU(3)_L\times U(1)_X$ is broken by a single $SU(3)_L$
triplet Higgs VEV \cite{liung}.

In addition to the gauge boson loop contributions, additional scalars may
also induce a $b\to s\gamma$ dipole transition.  The reduction of the
minimal Higgs sector of the 3-3-1 model has been considered in
\cite{liung}.  Three $SU(3)_L$ triplet and one sextet Higgs fields are
required to break the symmetries and generate all fermion masses.  While the
general Higgs structure is quite complicated (and includes flavor changing
neutral Higgs interactions), we make the assumption that
only interactions proportional to the (large) top Yukawa coupling are
important.  Along this line, we note that the minimal 3-3-1 Higgs sector
reduces to a three-Higgs-doublet SM with additional fields carrying lepton
number.  In the quark sector, this reduces further into a two-Higgs-doublet
model with the added feature that the couplings of the two
Higgs doublets to the third family are interchanged compared to the first
two families.

Assuming an approximately diagonal family structure, since
only loops involving the top quark are important, we only consider the third
family Higgs boson couplings.  In this case, the scalar contribution is
equivalent
to that of an ordinary two-Higgs-doublet model (at this level of
approximation), with the well known result
\begin{eqnarray}
C_7^{\rm 2HD}(M_W)&=&-{1\over6}V_{ts}^*V_{tb}^{\vphantom{*}}\left[
\left({2\over3}f_1(y_t)-f_2(y_t)\right)\cot^2\beta
-\left({2\over3}f_3(y_t)-f_4(y_t)\right)\right]\nonumber\\
C_8^{\rm 2HD}(M_W)&=&-{1\over6}V_{ts}^*V_{tb}^{\vphantom{*}}\left[
f_1(y_t)\cot^2\beta-f_3(y_t)\right]\ ,
\label{eq:hloop}
\end{eqnarray}
where
\begin{eqnarray}
f_3(y)&=&{3y(-y+3)\over2(y-1)^2}-{3y\over(y-1)^3}\ln y\nonumber\\
f_4(y)&=&{3y(y+1)\over2(y-1)^2}-{3y^2\over(y-1)^3}\ln y\ .
\end{eqnarray}
In this case, $y_t=m_t^2/M_{H^+}^2$ and $\tan\beta=v_2/v_1$ where $v_2$
gives rise to $m_t$.

The complete electroweak contribution to $b\to s\gamma$ in the 3-3-1 model
is simply a sum of the individual contributions given in
Eqns.~(\ref{eq:wloop}), (\ref{eq:yloop}), (\ref{eq:zploop}) and
(\ref{eq:hloop})
\begin{equation}
C_i(M_W) = C_i^{SM}(M_W) + C_i^{\rm 2HD}(M_W) + {M_W^2\over M_Y^2}
[C_i^Y(M_Y)+C_i^{Z'}(M_Y)]\ .
\end{equation}
This shows explicitly that the 3-3-1 contributions are suppressed by the
higher dilepton mass scale.  Nevertheless, as shown in the next section,
the 3-3-1 coefficient $C_7^Y(M_Y)$ is quite large because of the family
non-universality.  This and the upper bound on $M_Y$ ensure that the new
3-3-1 effects are generally of the same order as that of the SM and cannot
be ignored.

\section{The effective Hamiltonian and QCD corrections}

In the SM, the QCD corrections to $b\to s\gamma$ soften the GIM
mechanism to yield a logarithmic GIM cancellation.  This effect leads to an
enhancement of the $b\to s\gamma$ rate of about a factor of three.  For
the 3-3-1 model, however, the generalized GIM mechanism present in
dilepton exchange diagrams, Eqn.~(\ref{eq:yloop}), is imperfect, even in the
absence of QCD corrections.  Thus in this case, additional QCD corrections
are not expected to dominate the 3-3-1 contribution to $b\to s\gamma$.
Nevertheless, we include them here for completeness.

We deal with the QCD corrections using the standard technique of integrating
out the heavy degrees of freedom at each scale, using the renormalization
group approach with effective hamiltonians.  We work with three scales,
$m_b$, $M_W$ and $M_Y$, where the latter two can also be thought of as the
electroweak and the 3-3-1 scales respectively.  Starting at $M_Y$, we first
integrate out the 3-3-1 degrees of freedom, and then at $M_W$ integrate out
the $W$ and top simultaneously%
\footnote{In principle, there are additional corrections arising from QCD
running between $m_t$ and $M_W$ \cite{cho,adel}.  However such corrections
are presently dominated by the uncertainty arising from QCD scale
dependence in the leading order calculation and may be ignored.}.
Since the running between $M_W$
and $m_b$ has been extensively studied, we use the well known results of the
leading order calculation \cite{grinstein,misiak,ciuchini} and generalize
them to take into account additional operators present in the 3-3-1 case.

\subsection{The effective Hamiltonian for $M_W\le\mu\le M_Y$}
Below $M_Y$, we integrate out both dilepton and $Z'$ loops, yielding an
effective Hamiltonian
\begin{equation}
{\cal H}^\tto_{\rm eff}=-2\sqrt{2}G_F{M_W^2\over M_Y^2}
v_{3s}^*v_{3b}^{\vphantom{*}}\sum_i C_i^\tto(\mu)O^\tto_i(\mu)\ ,
\label{eq:heff331}
\end{equation}
where the set of operators $O^\tto_i$ consist of both four-Fermi and
penguin
operators relevant to the flavor changing $\Delta B = -\Delta S = -1$
interaction.  Since the 3-3-1 model has additional FCNC interactions, we
must use the extended operator basis
\begin{eqnarray}
O_1^t&=&(\overline{s}_{L\alpha}\gamma_\mu t_L^\beta)
	(\overline{t}_{L\beta}\gamma^\mu b_L^\alpha)\nonumber\\
O_2^t&=&(\overline{s}_L\gamma_\mu t_L)(\overline{t}_L\gamma^\mu b_L)
\nonumber\\
O_1^b&=&(\overline{s}_L\gamma_\mu b_L)(\overline{b}_L\gamma^\mu b_L)
\nonumber\\
O_3&=&(\overline{s}_L\gamma_\mu b_L)\sum_q(\overline{q}_L\gamma^\mu q_L)
\nonumber\\
O_4&=&(\overline{s}_{L\alpha}\gamma_\mu b_L^\beta)
	\sum_q(\overline{q}_{L\beta}\gamma^\mu q_L^\alpha)\nonumber\\
O_5&=&(\overline{s}_L\gamma_\mu b_L)\sum_q(\overline{q}_R\gamma^\mu q_R)
\nonumber\\
O_6&=&(\overline{s}_{L\alpha}\gamma_\mu b_L^\beta)
	\sum_q(\overline{q}_{R\beta}\gamma^\mu q_R^\alpha)\nonumber\\
O_5^Q&=&(\overline{s}_L\gamma_\mu b_L)
	\sum_q e_q(\overline{q}_R\gamma^\mu q_R)\nonumber\\
O_6^Q&=&(\overline{s}_{L\alpha}\gamma_\mu b_L^\beta)
	\sum_q e_q(\overline{q}_{R\beta}\gamma^\mu q_R^\alpha)\nonumber\\
O_7&=&{e\over16\pi^2}m_b(\overline{s}_L\sigma_{\mu\nu}b_R)F^{\mu\nu}
\nonumber\\
O_8&=&{g_3\over16\pi^2}m_b(\overline{s}_L\sigma_{\mu\nu}T^ab_R)G^{a\mu\nu}\ .
\label{eq:oper}
\end{eqnarray}
Color non-singlet channels are indicated explicitly via the color indices
$\alpha$ and $\beta$.  Since we are effectively above $m_t$, the sums are
over all six quarks.

The Wilson coefficients, $C_i^\tto(M_Y)$, are given by the matching
conditions at the 3-3-1 scale.  Integrating out the dileptons and exotic
quarks gives rise to the penguin operators as shown in the previous section.
Integrating out the $Z'$ gives rise to effective four-Fermi in addition to
the penguin operators.  To leading order, the non-zero Wilson coefficients
are
\begin{eqnarray}
C_{1t}^\tto(M_Y)&=&{2\over3}\rho_\tto\nonumber\\
C_{1b}^\tto(M_Y)&=&{2\over3}\rho_\tto\nonumber\\
C_3^\tto(M_Y)&=&{1\over3}\rho_\tto{2s^2-1\over c^2}\nonumber\\
C_{5Q}^\tto(M_Y)&=&2\rho_\tto{s^2\over c^2}\nonumber\\
C_7^\tto(M_Y)&=&{9\over4}-{8\over27}\rho_\tto{s^2\over c^2}
-\left[{2\over3}f_1(x_S)-{1\over2}f_2(x_S)\right]
+\left[-{5\over6}f_1(x_T)+f_2(x_T)\right]\nonumber\\
C_8^\tto(M_Y)&=&{8\over9}\rho_\tto{s^2\over c^2}
+{1\over2}\left[f_1(x_S)-f_1(x_T)\right]\ ,
\label{eq:m331}
\end{eqnarray}
where $x_S=m_S^2/M_Y^2$ and $x_T=m_T^2/M_Y^2$.  We have assumed negligible
mixing to the first family so that $v_{2s}^*v_{2b}^{\vphantom{*}}\approx
- v_{3s}^*v_{3b}^{\vphantom{*}}$.
This is also the reason why the exotic $D$ quark does not appear.  Although
we take $\rho_\tto=3/4$ for our numerical results, it is shown
explicitly above to indicate that those terms are due to $Z'$ exchange.

The $Z'$ induced terms arise where one vertex gives the $b\to s$ flavor
changing interaction and the other vertex is flavor conserving.  However,
since the left handed $Z'$ interaction is non-universal, it singles out one
of the families for special treatment.  Up to small mixing (which we
ignore), it must be the third family \cite{liu,ppfcnc}, which is why the
third family operators $O_1^t$ and $O_1^b$ are singled out.
The other new four-Fermi operators, $O_2^t$, $O_5^Q$ and $O_6^Q$, must then
be included to account for operator mixing as well as the right
handed $Z'$ vertex.  It is straightforward to extend the results presented in
\cite{ciuchini} to determine the anomalous dimension matrix corresponding
to the mixing of the operators in Eqn.~(\ref{eq:oper}).  The results are shown
in the Appendix.

In order to examine the significance of the additional 3-3-1 contributions
to $b\to s\gamma$, we show values for the initial Wilson coefficient
$C_7^\tto(M_Y)$ in Fig.~\ref{fig:3}.  From the figure, it is obvious that
the generalized GIM mechanism is quite different from the ordinary case.
Instead of vanishing as in the usual case, the Wilson coefficient takes on
its largest values when the exotic quarks are relatively light and
degenerate in mass.  Since the functions $f_1$ and $f_2$ are bounded by
$0\le f_1(x)\le 1/4$ and $-1/2\le f_2(x)\le 0$, we find the limits
\begin{eqnarray}
1.06\le C_7^\tto(M_Y)\le2.18\nonumber\\
0.076\le C_8^\tto(M_Y)\le0.326\ ,
\label{eq:limits}
\end{eqnarray}
although for realistic values of $m_S$ and $m_T$, we expect a more limited
range, $1.3\alt C_7^\tto(M_Y)\alt 2.0$ as indicated by the figure.  Note
that this may be contrasted with the SM value $C_7^{\rm SM}(M_W)
\approx -0.20$ for $m_t=175$~GeV.  After accounting for the additional
$M_W^2/M_Y^2$ factor in Eqn.~(\ref{eq:heff331}) that arises from the difference
in mass scales, we see that the new 3-3-1 effects are comparable to that of
the SM.

\subsection{The matching conditions at $M_W$ and ${\cal H}_{\rm eff}$ for
$m_b\le\mu\le M_W$}
When we reach the electroweak scale, $M_W$, we further integrate out the top
and $W$.  In this case, we match the effective hamiltonian ${\cal
H}^\tto_{\rm eff}(M_W)$ onto a second one, ${\cal H}_{\rm eff}$, without
the top degrees of freedom.  At this stage, since we include the SM
contributions to $b\to s\gamma$, we choose the conventional form
\begin{equation}
{\cal H}_{\rm eff}=-2\sqrt{2}G_FV_{ts}^*V_{tb}^{\vphantom{*}}
\sum_i C_i(\mu)O(\mu)\ .
\end{equation}
However, since we have included a larger set of operators in
${\cal H}^\tto_{\rm eff}$, they must be retained when running to $m_b$.
Thus a complete set of operators below $M_W$ consist of those of
Eqn.~(\ref{eq:oper}), with the exception that $O_1^t$ and $O_2^t$ are replaced
by the conventional operators
\begin{eqnarray}
O_1&=&(\overline{s}_{L\alpha}\gamma_\mu c_L^\beta)
	(\overline{c}_{L\beta}\gamma^\mu b_L^\alpha)\nonumber\\
O_2&=&(\overline{s}_L\gamma_\mu c_L)(\overline{c}_L\gamma^\mu b_L)\ ,
\end{eqnarray}
and we only take five active quarks in $O_3$ through $O_6^Q$.

At $M_W$, the Wilson coefficients get contributions both from integrating
out the top and from matching onto ${\cal H}^\tto_{\rm eff}$.  We find
\begin{eqnarray}
C_1(M_W)&=&C_1^{\rm 2HDSM}(M_W)\nonumber\\
C_2(M_W)&=&C_2^{\rm 2HDSM}(M_W)\nonumber\\
C_i(M_W)&=&C_i^{\rm 2HDSM}(M_W)+\chi {M_W^2\over M_Y^2} C_i^\tto(M_W)
\qquad \hbox{(all other operators)}\ ,
\end{eqnarray}
where $\chi=v_{3s}^*v_{3b}^{\vphantom{*}}/V_{ts}^*V_{tb}^{\vphantom{*}}$ is the
ratio of 3-3-1 and SM mixing angles.  The SM coefficients include the
contributions from the two-Higgs-doublet model and are given by
\begin{eqnarray}
C_2^{\rm 2HDSM}(M_W)&=&1\nonumber\\
C_7^{\rm 2HDSM}(M_W)&=&-{1\over2}\left[{2\over3}f_1(x_t)-f_2(x_t)\right]
\nonumber\\
&&-{1\over6}\left[\left({2\over3}f_1(y_t)-f_2(y_t)\right)\cot^2\beta
-\left({2\over3}f_3(y_t)-f_4(y_t)\right))\right]\nonumber\\
C_8^{\rm 2HDSM}(M_W)&=&-{1\over2}f_1(x_t)\nonumber\\
&&-{1\over6}\left[f_1(y_t)\cot^2\beta-f_3(y_t)\right]\ .
\label{eq:msm}
\end{eqnarray}

\section{The $\lowercase{b}\to \lowercase{s}\gamma$ rate and limits on
3-3-1 physics}
Once the 3-3-1 and SM matching conditions, Eqns.~(\ref{eq:m331}) and
(\ref{eq:msm}), are given, it is straightforward to solve the
renormalization group equations to arrive at $C_7(\mu)$.  The $b\to s\gamma$
decay rate is then calculated in the ratio
\begin{equation}
{\Gamma(b\to s\gamma)\over\Gamma(b\to ce\overline{\nu}_e)}
={|V_{ts}^*V_{tb}^{\vphantom{*}}|^2\over|V_{cb}|^2}{6\alpha\over\pi I(z)}
|C_7(\mu)|^2\ ,
\label{eq:rate}
\end{equation}
where $z=m_c/m_b$ and $I(z)=1-8z^2+8z^6-z^8-24z^4\ln z$ is the phase space
factor for the charged current decay; $I(z)=0.485\pm0.028$ for
$z=0.316\pm0.013$ \cite{ruckl}.  Following \cite{buras}, we do not include
the $O(\alpha_3)$ corrections to $b\to ce\overline{\nu}_e$ at this order.
In the absence of a complete next to leading order calculation, when
necessary we vary the
renormalization scale $\mu$ by a factor of two around $m_b$ to estimate the
effects of the QCD scale ambiguity \cite{buras}.  The resulting large $\mu$
dependence of the leading order calculation is the dominant theoretical
uncertainty in the predicted $b\to s\gamma$ decay rate.

Using the current value for the charged current decay mode, ${\rm BR}(b\to
ce\overline{\nu}_e)=(0.104\pm0.004)$ \cite{pdb}, Eqn.~(\ref{eq:rate}) may
be rewritten as
\begin{equation}
{\rm BR}(b\to s\gamma)=(2.84\pm0.23)\times 10^{-3}|C_7(\mu)|^2\ ,
\end{equation}
so that the current CLEO limits, Eqn.~(\ref{eq:cleo}), correspond to
\begin{equation}
0.18\le|C_7(\mu)|\le0.38\qquad\hbox{(CLEO)}\ .
\end{equation}
{}From now on we focus on $C_7(\mu)$ instead of the branching ratio since
the SM $W$--$t$, $H^+$--$t$ and new 3-3-1 contributions may simply be added
together in the amplitude (this will always be true, even with higher order
QCD corrections, as long as we work only at the one-electroweak-loop order).

The 3-3-1 model introduces many new parameters into the $b\to s\gamma$
calculation.  Explicitly, we write
\begin{eqnarray}
C_7(\mu)&=&C_7(\mu;m_t,M_{H^+},\tan\beta;m_S,m_T,M_Y,\chi)\nonumber\\
&=&C_7^{\rm 2HDSM}(\mu;m_t,M_{H^+},\tan\beta)
+\Delta C_7(\mu;m_S,m_T,M_Y,\chi)\ ,
\end{eqnarray}
where in the second line we have separated out the two-Higgs-doublet SM and
new 3-3-1 contributions.  The SM prediction for $C_7(m_b)$ is shown by the
solid line in Fig.~\ref{fig:4} along with the CLEO bounds.  We have taken
$\alpha_3(M_Z)=0.120$ and $m_b=5$~GeV.  As can be seen,
the present experimental data is consistent with the unadorned SM.  When the
charged Higgs loop is included, it always contributes with the same sign and
hence can only increase the predicted branching ratio.  As an example, we
plot $C_7(m_b)$ for several values of the charged Higgs mass in
Fig.~\ref{fig:4} (dotted lines) when $\tan\beta=3$.  The two-Higgs-doublet
results are fairly insensitive to $\tan\beta$ provided $\tan\beta\agt1$.
For small $\tan\beta$, on the other hand, the charged Higgs contribution is
enhanced and often falls outside the CLEO bound.

The effect of the 3-3-1 contributions, $\Delta C_7(m_b)$, may be estimated
from the renormalization group analysis as
\begin{equation}
\Delta C_7(m_b)\approx \chi {M_W^2\over M_Y^2}\left[
0.063 + 0.59 C_7^\tto(M_Y) + 0.11 C_8^\tto(M_Y)\right]\ ,
\end{equation}
using 400~GeV as the 3-3-1 scale.  Since both Wilson coefficients are
bounded according to Eqn.~(\ref{eq:limits}), we find the size of the new
3-3-1 effects to be
\begin{equation}
\Delta C_7(m_b)\approx\chi {M_W^2\over M_Y^2}(0.7 \hbox{ -- }1.4)\ .
\label{eq:approx331}
\end{equation}
This range is given by the dotted lines in Fig.~\ref{fig:5}, assuming
$\chi=1$.  The actual predictions for degenerate exotic quarks are given by
the solid lines in the figure.

While the mixing parameter, $\chi=v_{3s}^*v_{3b}^{\vphantom{*}}/
V_{ts}^*V_{tb}^{\vphantom{*}}$ is in
principle undetermined, we expect both the numerator and denominator to be
comparable, giving a value $|\chi|\sim O(1)$.  Unrealistic cases of
vanishing mixing in the down quark sector ($V_L=1$) and up quark sector
($U_L=1$) give $\chi=0$ and $1$ respectively.  In order to put an
experimental limit on $\chi$, we may use the neutral meson mixing data to
restrict the product of 3-3-1 mixing angles to be
$|v_{3s}^*v_{3b}^{\vphantom{*}}|\le 0.25$
at 90\% C.L.~\cite{liu}.  This gives the bound $|\chi|\le 8.3$, although
values near the upper bound may be somewhat unnatural from a theoretical
point of view.

As in the two-Higgs-doublet model, the additional contribution from the
3-3-1 model, $\Delta C_7(\mu)$, always enters with the same sign.  However,
unlike the charged Higgs contribution which always {\it increases} the
$b\to s \gamma$ rate, the effect of the new 3-3-1 contribution depends on the
quark mixing parameter $\chi$.  From Eqn.~(\ref{eq:approx331}) we see that
$\chi>0$ ($<0$) leads to a suppression (enhancement) of the overall
$b\to s\gamma$ decay rate.  In general, when $\chi$ is complex, the results
lie somewhere in the middle.  This feature seems to be shared with other
models incorporating two Higgs doublets.  Namely, while the charged Higgs
loop alone always increases the $b\to s \gamma$ decay rate,  the additional
new particles, whether superpartners in the SUSY case or dilepton gauge
bosons in the 3-3-1 model, contribute with arbitrary sign and may compensate
for the increase arising from the Higgs loop.  This is also the case for
radiative corrections to the $Z\to b\overline{b}$ vertex.

In order to examine how the new 3-3-1 physics weakens the $b\to s \gamma$
limits on the pure two-Higgs-doublet model, we plot the allowed region in
the $\tan\beta$--$M_{H^+}$ parameter space in Fig.~\ref{fig:6}.  In this
case we have fixed the top quark mass to be 175 GeV.  The solid line
corresponds to the two-Higgs-doublet model and shows that
$M_{H^+}>290$~GeV for the case $m_t=175$~GeV.  In arriving at this limit we
have estimated the theoretical uncertainty by varying the QCD scale $\mu$
from $m_b/2$ to $2 m_b$.  As noted previously, the
limits are insensitive to $\tan\beta$ when $\tan\beta\agt1$.  Inclusion of
3-3-1 physics with $|\chi|\le1$ lowers this bound as indicated by the dotted
lines in the figure.  For a dilepton gauge boson mass $M_Y=300$~GeV, the
corresponding limit on the charged Higgs mass is weakened to
$M_{H^+}>120$~GeV for light exotic quarks.  The 3-3-1 limits correspond to
real positive $\chi$, which is the region of maximum cancellation between
the charged Higgs and dilepton gauge boson loops.  Larger values of $\chi$
naturally weaken the limits further.  However the further we are below the
two-Higgs-doublet limit (solid line), the more tuning is required between
the two-Higgs-doublet and 3-3-1 parameters to achieve large cancellations.
Thus from a naturalness point of view, we expect the charged Higgs mass in
the 3-3-1 model to be no lighter than $\sim120$~GeV.

\section{Conclusion}

Because it is a loop process, the FCNC decay $b\to s \gamma$ presents an
interesting test of both SM and new physics.  In the SM, this process is
GIM suppressed and proceeds through a heavy top quark.  While the
theoretical calculation suffers from large uncertainties due to unknown next
to leading order QCD corrections, it is in excellent agreement with the
current CLEO data \cite{cleo}.  Nevertheless, at the present level, the
results are not yet sensitive to $m_t$ in the SM, as may be seen from
Fig.~\ref{fig:4}.  Future work, on both the theoretical and
experimental side, may bring the uncertainties down to the point where
$b\to s \gamma$ would become more sensitive to both $m_t$ and new physics.

In anticipation of such future improvements, we may estimate the size of
the contribution of new physics to $b\to s\gamma$.  Compared to the SM
value, $C_7(m_b)\sim -0.3$, new physics at a scale $M_{\rm new}$ is expected
to contribute roughly to the $b\to s\gamma$ vertex as
$|\delta C_7(m_b)|\sim (M_W^2/M_{\rm new}^2) (\Delta M_Q^2/M_{\rm new}^2)$
where $\Delta M_Q$ is a typical mass splitting between the new fermions in
the loop and arises via a generalized GIM mechanism.  Thus in general
(assuming the absence of tree level FCNCs in the extended model) new physics is
suppressed by both the heavier mass scale and a generalized GIM mechanism
and is hence dominated by the larger SM contribution.

In order to evade this conclusion, we need to either have $M_{\rm new}
\approx M_W$ or somehow avoid the generalized GIM cancellation.  An example
of the former case is the two-Higgs-doublet model where a light charged
Higgs particle may be eliminated by the current CLEO data.  As an example
of the latter case, we have performed a detailed calculation of $b\to
s\gamma$ in the 3-3-1 model.  Due to the different representation of the
third quark family, the generalized GIM cancellation is imperfect and
$|\Delta C_7(m_b)| \sim M_W^2/M_Y^2$ is non-vanishing even for degenerate
exotic quark masses.  Compared to the two-Higgs-doublet model where
$M_{H^+}>290$~GeV, this limit is weakened to $M_{H^+}\agt120$~GeV in the
full 3-3-1 model.

Although the 3-3-1 model appears tightly constrained both by FCNC limits and
by the large $U(1)_X$ coupling \cite{ppframpton,daniel,willen}, the model
survives the test of $b\to s\gamma$, at least up to the current level of
precision.  As experimental evidence for a heavy top continues to build,
the curious feature of a different third generation in the 3-3-1 model takes
on more significance.  Hence we look forward with anticipation to what
future experiments in the $B$ system will bring.

\bigskip
This work was supported in part by the National Science Foundation under Grant
No.~PHY-9411543, and by the Department of Energy under Grant
No.~DE-FG05-85ER-40219.  JTL wishes to thank D.~Ng for enlightning
discussions in the initial stages of this work.

\appendix
\section*{The anomalous dimension matrices}
At leading order, the renormalization group equation for the Wilson
coefficients $\vec C(\mu)$ is
\begin{equation}
\mu{d\over d\mu}\vec C(\mu)=\left[\gamma^T(\mu)\right]\vec C(\mu)\ ,
\label{eq:rge}
\end{equation}
where
\begin{equation}
\gamma(\mu)={\alpha_3(\mu)\over2\pi}\gamma^0\ ,
\end{equation}
and $\alpha_3(\mu)\equiv g_3^2(\mu)/4\pi$ satisfies the $\beta$-function
equation
\begin{equation}
\mu{d\over d\mu}\alpha_3^{-1}(\mu)=-{b\over2\pi}\ .
\end{equation}
Here $b=-11N/3+2f/3$ is the one-loop QCD beta function coefficient for $f$
quarks ($b_{(6)}=-7$ and $b_{(5)}=-23/3$).  Eqn.~(\ref{eq:rge}) is exactly
solved by
\begin{equation}
\vec C(\mu)=V\left[\eta^{-\gamma_D^0/b}\right]V^{-1}\vec C(M)\ ,
\end{equation}
where $\eta=\alpha_3(M)/\alpha_3(\mu)$.  For consistency at this order, the
running of $\alpha_3(\mu)$ is only calculated to one loop.  In
this notation, $\gamma_D^0=V^{-1}\gamma^{0T}V$ is the diagonal matrix of
eigenvalues of $\gamma^0$.

For the extended operator basis of Eqn.~(\ref{eq:oper}), we write the scheme
independent anomalous dimension matrix in the form
\begin{equation}
\gamma^0 =\pmatrix{\gamma_{44}&\gamma_{4P}\cr
0&\gamma_{PP}}\ .
\end{equation}
Then, for $u$ up and $d$ down type quarks, we find
\begin{equation}
\gamma_{44}=\pmatrix{-3\over N&3&0&0&0&0&0&0&0\cr
3&-3\over N&0&-{1\over3N}&1\over3&-{1\over3N}&1\over3&0&0\cr
0&0&3-{3\over N}&-{1\over3N}&1\over3&-{1\over3N}&1\over3&0&0\cr
0&0&0&-{3\over N}-{2\over3N}&3+{2\over3}&-{2\over3N}&2\over3&0&0\cr
0&0&0&3-{f\over3N}&-{3\over N}+{f\over3}&-{f\over3N}&f\over3&0&0\cr
0&0&0&0&0&3\over N&-3&0&0\cr
0&0&0&-{f\over3N}&f\over3&-{f\over3N}&-6C_2+{f\over3}&0&0\cr
0&0&0&0&0&0&0&3\over N&-3\cr
0&0&0&-{\overline{f}\over3N}&\overline{f}\over3&-{\overline{f}\over3N}&
\overline{f}\over3&0&-6C_2}\ ,
\end{equation}
where $f=u+d$ and $\overline{f}=e_u u + e_d d=(2u-d)/3$ and
$C_2=(N^2-1)/2N$.  The mixing of the penguin operators are given by
\begin{equation}
\gamma_{PP}=\pmatrix{4C_2&0\cr-{1\over3}(4C_2)&8C_2-2N}\ .
\end{equation}
For $\gamma_{4P}$, we use the scheme independent formalism of \cite{ciuchini}
(which, at this level, is equivalent to using the 't Hooft--Veltman
regularization scheme).  The result is
\begin{equation}
\gamma_{4P}=\pmatrix{0&3S_2\cr
(3 e_u+{2\over9}e_d)C_2&{29\over9}C_2-N\cr
(3+{2\over9})e_dC_2&3S_2+{29\over9}C_2-N\cr
2(3+{2\over9})e_dC_2&3fS_2+2({29\over9}C_2-N)\cr
(3\overline{f}+{2\over9}fe_d)C_2&6S_2+f({29\over9}C_2-N)\cr
-4e_dC_2&-3fS_2-4C_2+N\cr
(-3\overline{f}+{2\over9}fe_d)C_2&-4S_2+f(-{25\over9}C_2+{N\over2})\cr
-4e_d^2C_2&-3\overline{f}S_2+e_d(-4C_2+N)\cr
(-3(u e_u^2+d e_d^2)+{2\over9}\overline{f}e_d)C_2&
-4e_dS_2+\overline{f}(-{25\over9}C_2+{N\over2})}\ ,
\end{equation}
where $S_2=1/2$.

For $M_W\le\mu\le M_Y$, we set $N=3$ and six quarks are active ($u=d=3$).
In this case the explicit anomalous dimension matrix is
\begin{equation}
\gamma^0_{(6)}=\pmatrix{-1&3&0&0&0&0&0&0&0&0&{3\over2}\cr
3&-1&0&-{1\over9}&1\over3&-{1\over9}&1\over3&0&0&208\over81&35\over27\cr
0&0&2&-{1\over9}&1\over3&-{1\over9}&1\over3&0&0&-{116\over81}&151\over54\cr
0&0&0&-{11\over9}&11\over3&-{2\over9}&2\over3&0&0&-{232\over81}&313\over27\cr
0&0&0&7\over3&1&-{2\over3}&2&0&0&92\over27&97\over9\cr
0&0&0&0&0&1&-3&0&0&16\over9&-{34\over3}\cr
0&0&0&-{2\over3}&2&-{2\over3}&-6&0&0&-{124\over27}&-{137\over9}\cr
0&0&0&0&0&0&0&1&-3&-{16\over27}&-{13\over18}\cr
0&0&0&-{1\over9}&1\over3&-{1\over9}&1\over3&0&-8&-{548\over81}&-{83\over54}\cr
0&0&0&0&0&0&0&0&0&16\over3&0\cr
0&0&0&0&0&0&0&0&0&-{16\over9}&14\over3}\ .
\end{equation}
Below $M_W$, the top quark is integrated out, and we are left with a five
quark anomalous dimension matrix
\begin{equation}
\gamma^0_{(5)}=\pmatrix{-1&3&0&0&0&0&0&0&0&0&{3\over2}\cr
3&-1&0&-{1\over9}&1\over3&-{1\over9}&1\over3&0&0&208\over81&35\over27\cr
0&0&2&-{1\over9}&1\over3&-{1\over9}&1\over3&0&0&-{116\over81}&151\over54\cr
0&0&0&-{11\over9}&11\over3&-{2\over9}&2\over3&0&0&-{232\over81}&545\over54\cr
0&0&0&22\over9&2\over3&-{5\over9}&5\over3&0&0&68\over81&256\over27\cr
0&0&0&0&0&1&-3&0&0&16\over9&-{59\over6}\cr
0&0&0&-{5\over9}&5\over3&-{5\over9}&-{19\over3}&0&0&
-{148\over81}&-{703\over54}\cr
0&0&0&0&0&0&0&1&-3&-{16\over27}&5\over18\cr
0&0&0&-{1\over27}&1\over9&-{1\over27}&1\over9&0&-8&
-{1196\over243}&-{11\over162}\cr
0&0&0&0&0&0&0&0&0&16\over3&0\cr
0&0&0&0&0&0&0&0&0&-{16\over9}&14\over3}\ .
\end{equation}

\begin{figure}
\caption{The gauge boson contributions to $b\to s\gamma$.  In the 3-3-1
model, $V$ may be either $W$, $Y$ or $Z'$.}
\label{fig:1}
\end{figure}

\begin{figure}
\caption{Additional diagram needed to cancel divergences and restore gauge
invariance in the $b\to s\gamma$ vertex in the 3-3-1 model.}
\label{fig:2}
\end{figure}

\begin{figure}
\caption{Contours of constant $C_7^\tto(M_Y)$ in the $x_T$--$x_S$ plane.
Note that $C_7^\tto(M_Y)$ is everywhere positive and does not vanish on the
diagonal $x_T=x_S$.}
\label{fig:3}
\end{figure}

\begin{figure}
\caption{The Wilson coefficient $C_7(m_b)$ in the SM (solid line) and
two-Higgs-doublet model (dotted lines) plotted as a function of $m_t$.
For the two-Higgs-doublet model, we have taken $\tan\beta=3$.  The current
CLEO bounds are shown by the dashed lines.}
\label{fig:4}
\end{figure}

\begin{figure}
\caption{The 3-3-1 model contribution to $C_7(m_b)$ (with the mixing
parameter $\chi$ removed) plotted as a function
of dilepton gauge boson mass for degenerate exotic quark masses $m_Q=250$,
500, 750 and 1000 GeV (solid lines).  The dotted lines indicate the minimum
and maximum possible values arising from 3-3-1 physics.}
\label{fig:5}
\end{figure}

\begin{figure}
\caption{The allowed region in $\tan\beta$--$M_{H^+}$ parameter space for
the two-Higgs-doublet model (solid line) and the 3-3-1 model (dotted lines).
For each pair of dotted lines, the upper one corresponds to $M_Q=1000$~GeV
and the lower one to $M_Q=250$~GeV.  For the 3-3-1 model we have restricted
the ratio of mixing angles by $|\chi|\le1$.}
\label{fig:6}
\end{figure}


\newpage
\input psfig
\ \vskip 1in
\centerline{\psfig{figure=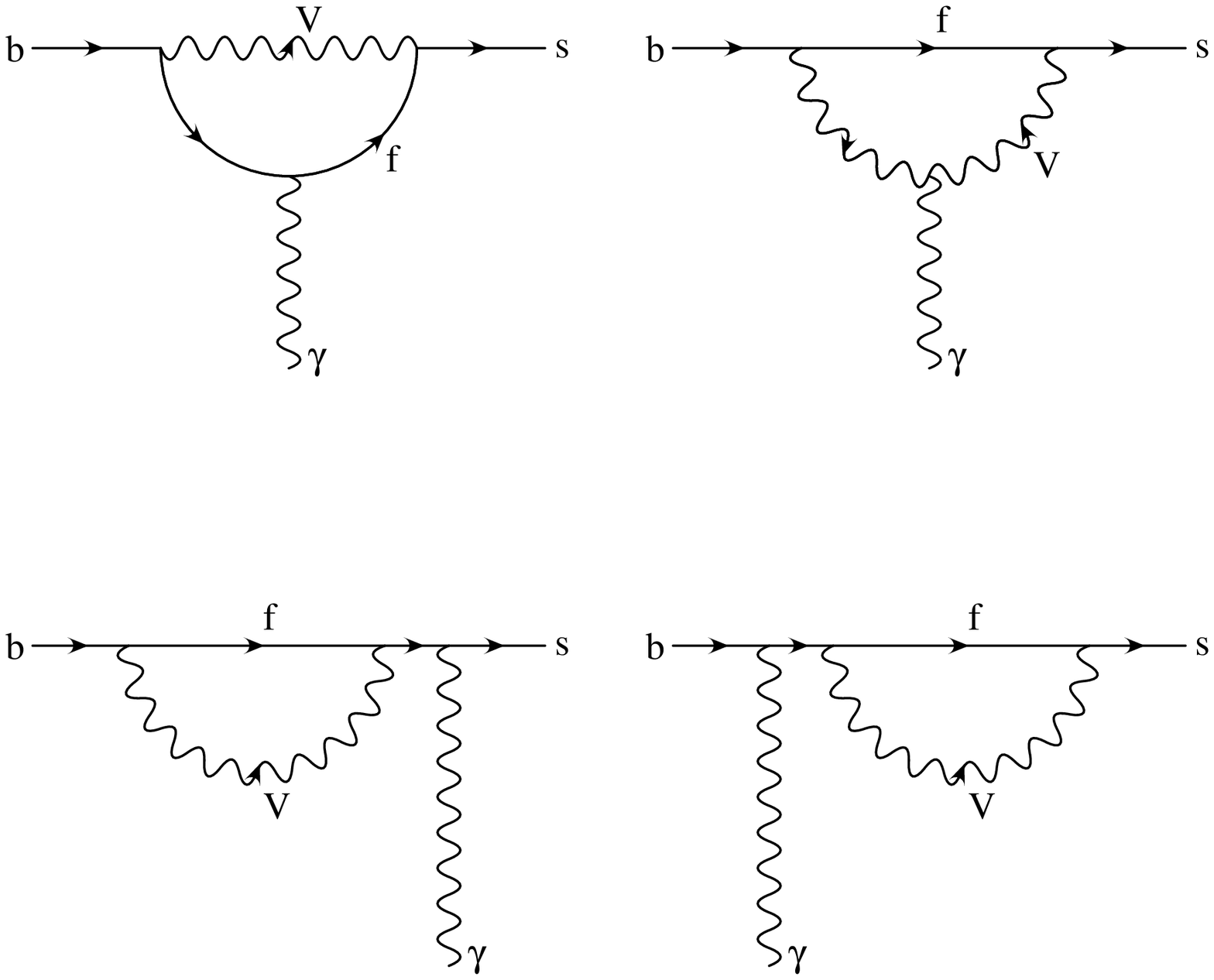}}
\vskip 1in
\centerline{\bf Figure 1}

\newpage
\ \vskip 1.5in
\centerline{\psfig{figure=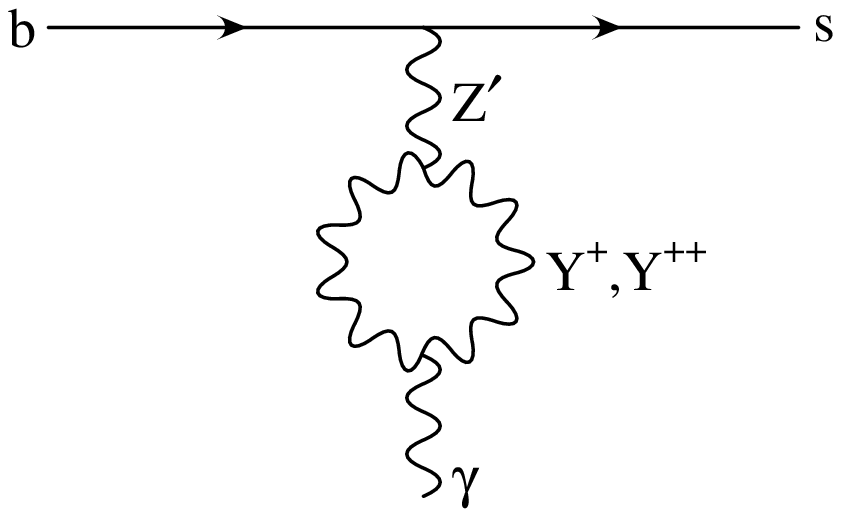}}
\vskip 1.5in
\centerline{\bf Figure 2}

\newpage
\centerline{\psfig{figure=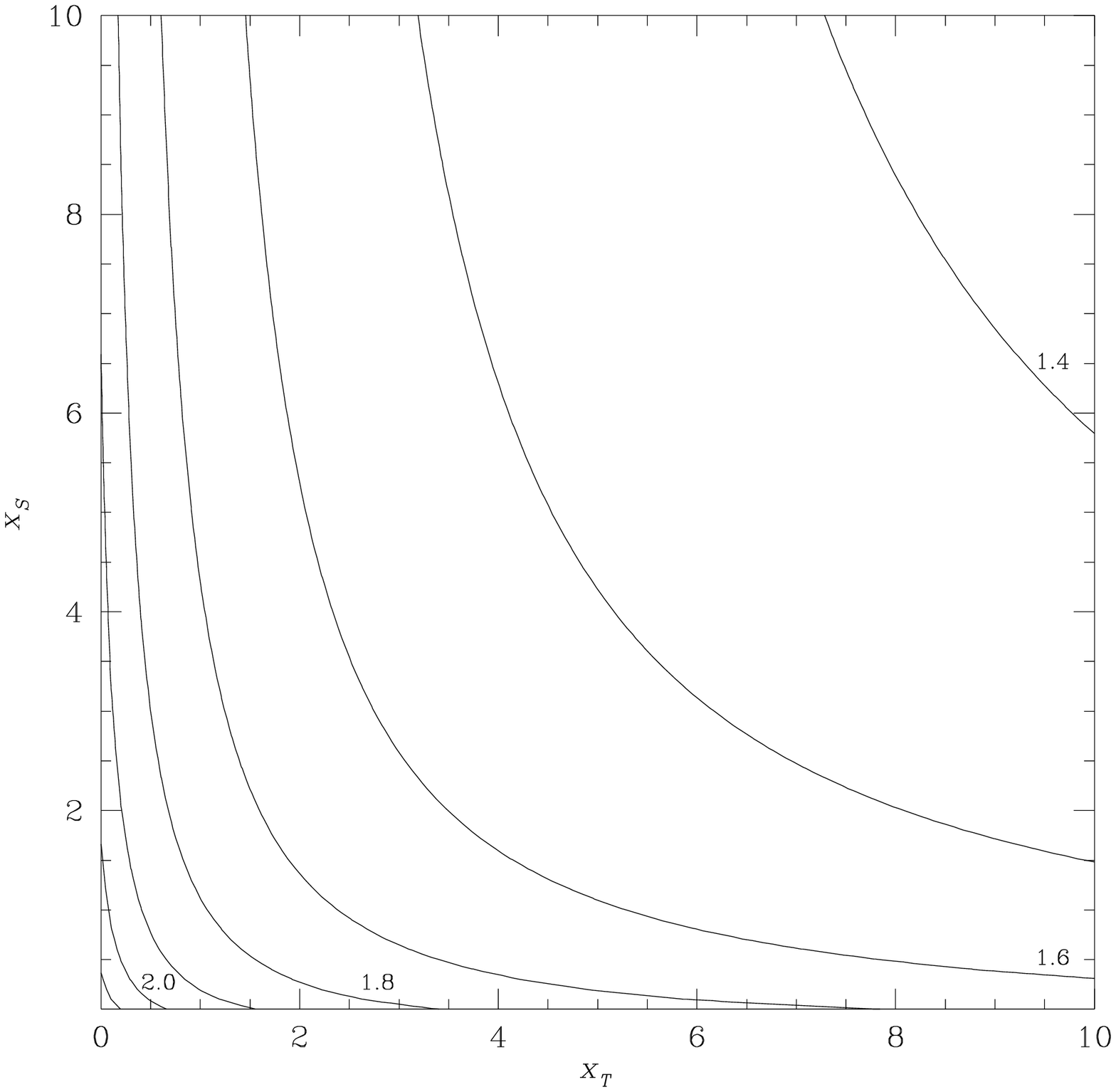}}
\centerline{\bf Figure 3}

\newpage
\centerline{\psfig{figure=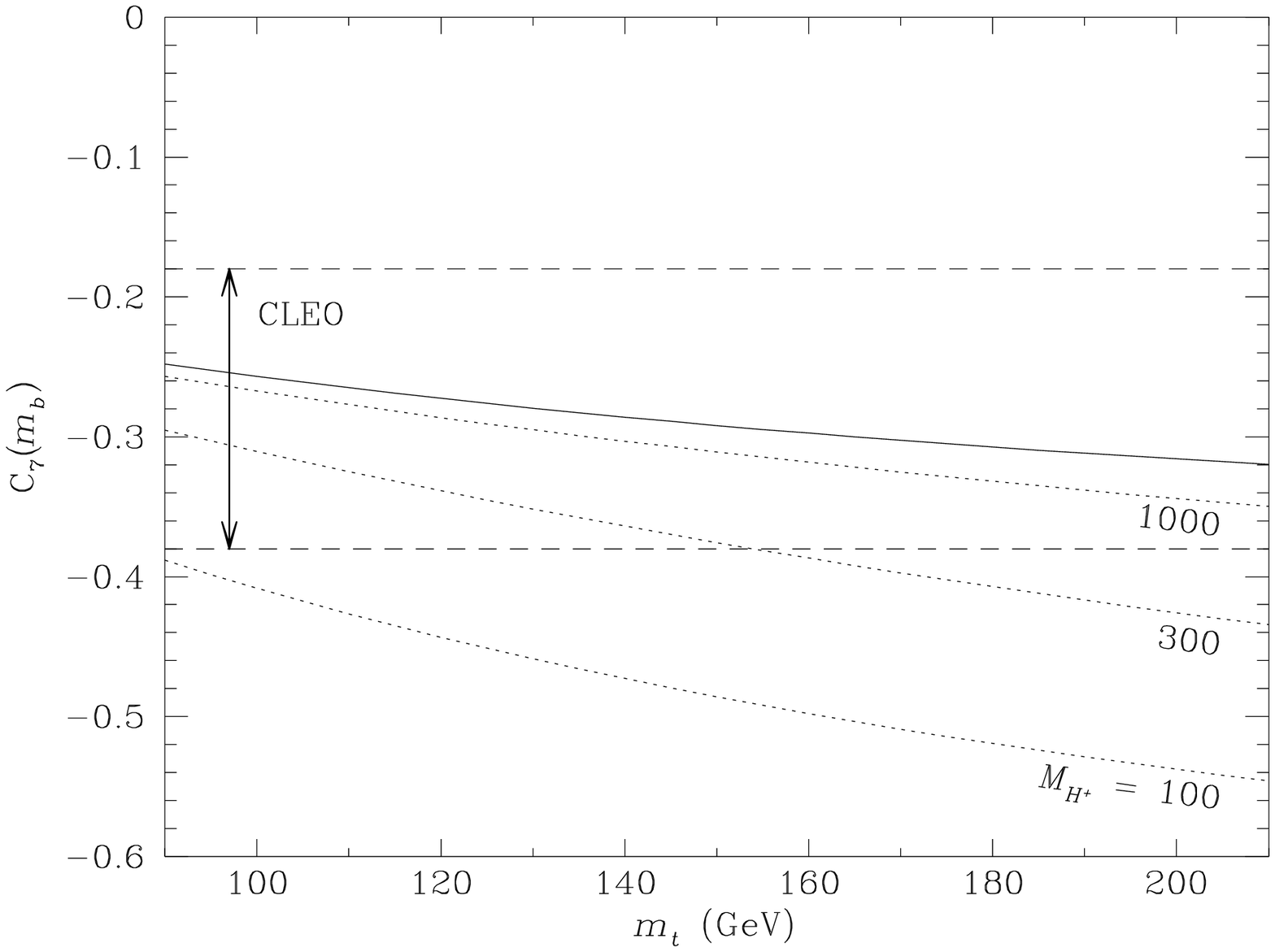}}
\vskip 1in
\centerline{\bf Figure 4}

\newpage
\centerline{\psfig{figure=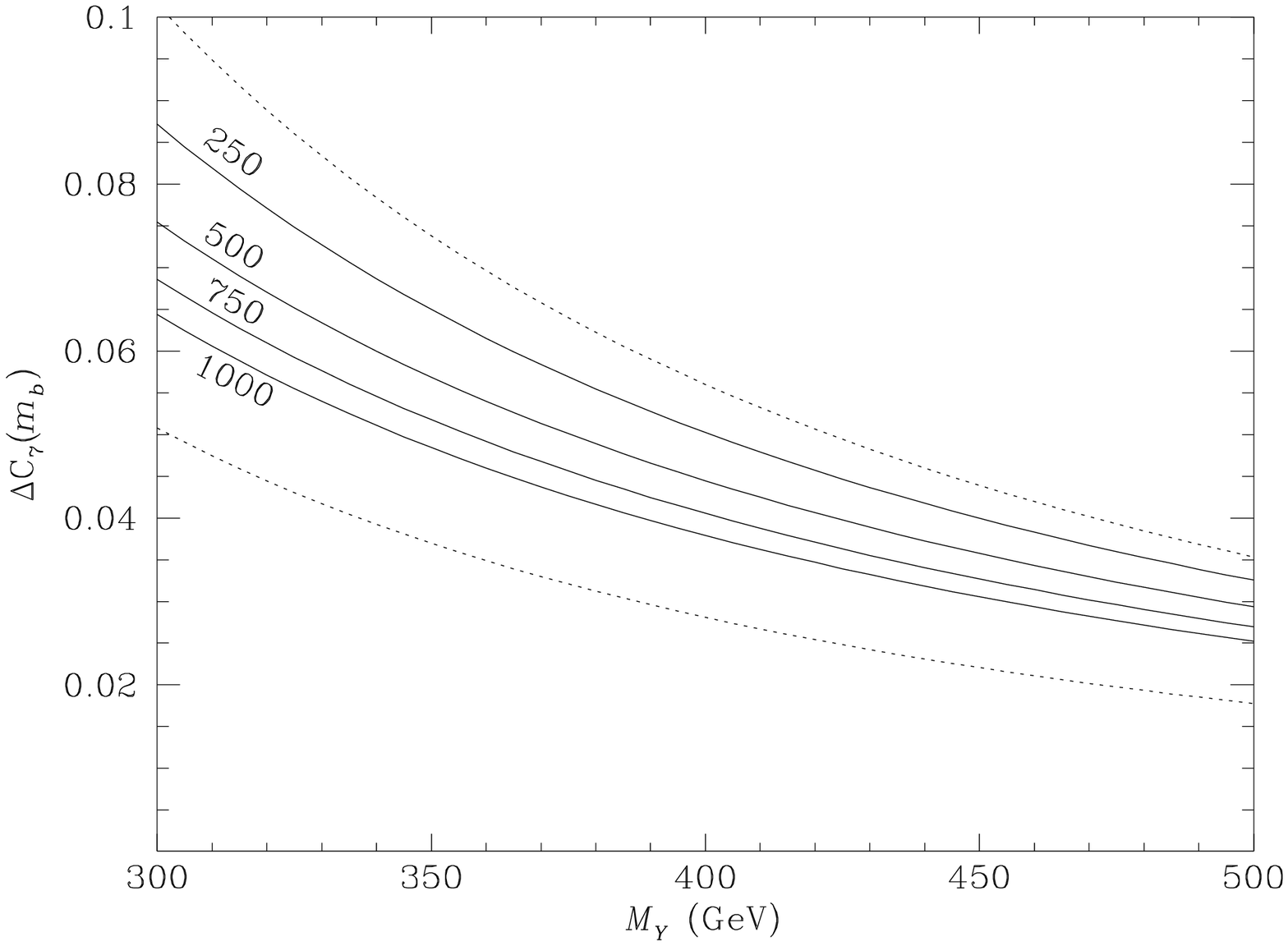}}
\vskip 1in
\centerline{\bf Figure 5}

\newpage
\centerline{\psfig{figure=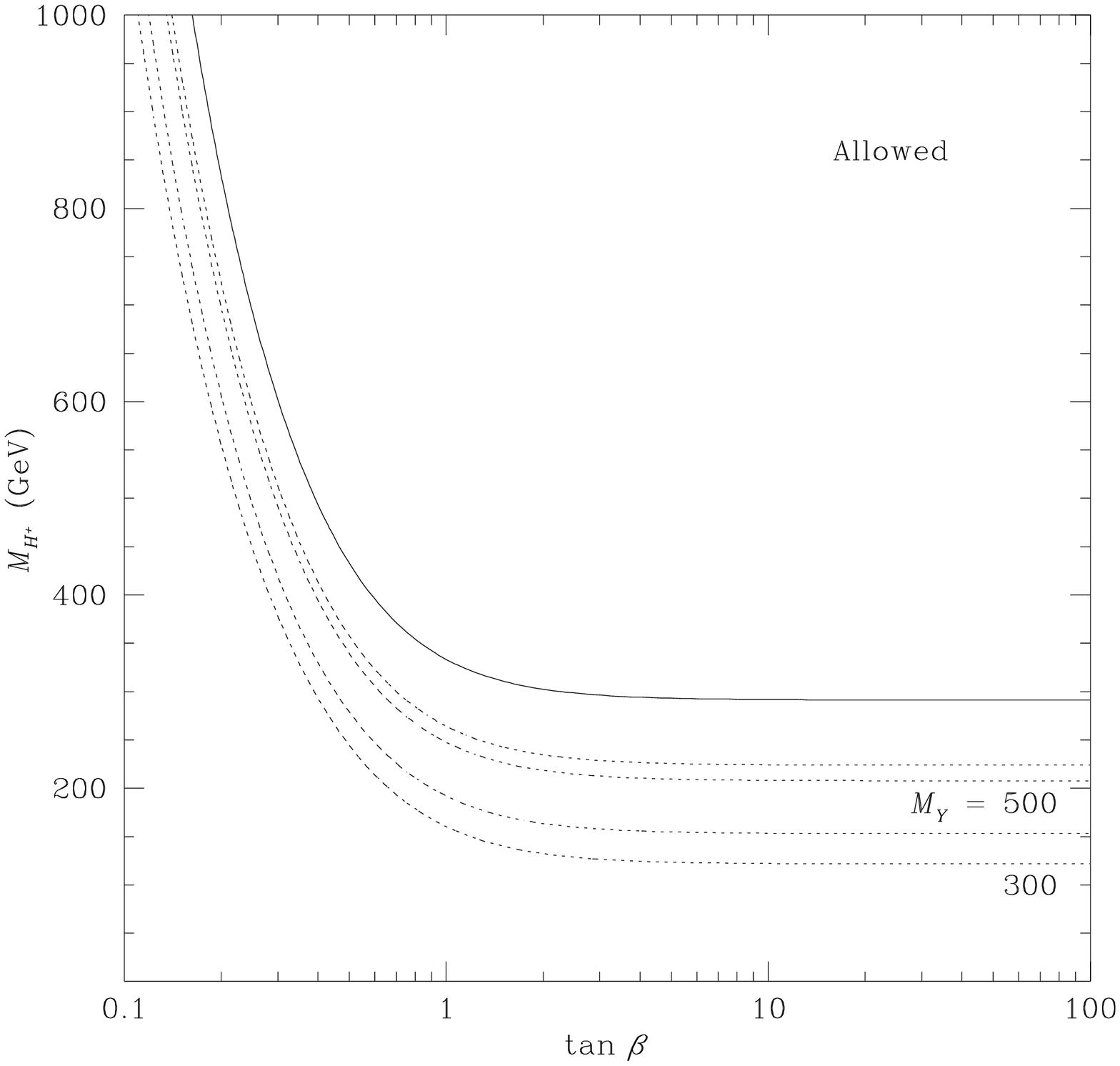}}
\centerline{\bf Figure 6}

\end{document}